\documentclass[aps,prb,twocolumn,footinbib,floatfix,superscriptaddress]{revtex4-1}  
\usepackage{CJK}
\usepackage{amsmath}
\usepackage{amsfonts}
\usepackage{bm}
\usepackage{soul}
\usepackage[caption=false,subrefformat=parens,labelformat=parens]{subfig}
\usepackage{color} 
\usepackage[table,dvipsnames]{xcolor}
\usepackage[colorlinks,linktocpage,bookmarks=false]{hyperref}
\usepackage{graphicx}% 

\newcommand\myshade{85}
\definecolor{myrulecolor}{RGB}{150,20,0}% define the color for the rules
\colorlet{mylinkcolor}{violet}
\colorlet{mycitecolor}{YellowOrange}
\colorlet{myurlcolor}{Aquamarine}
\hypersetup{
	linkcolor  = myurlcolor!\myshade!black,
	citecolor  = mycitecolor!\myshade!black,
	urlcolor   =  myrulecolor!\myshade!black,
}

\usepackage{graphicx}
\graphicspath{{./FIG/}}
\usepackage{ulem}
\usepackage{multirow}
\usepackage{braket}

\newcommand{\beq}{\begin{equation}}
\newcommand{\eeq}{\end{equation}}
\newcommand{\bea}{\begin{eqnarray}}
\newcommand{\eea}{\end{eqnarray}}

\newcommand{\matr}[1]{\mathbf{#1}}

\newcommand{\bfq}{\bm{q}}
\newcommand{\bfr}{\bm{r}}
\newcommand{\rr}{\bfr}
\newcommand{\qq}{\bm{q}}
\newcommand{\ud}{\mathrm{d}}
\newcommand{\bfu}{\bm{u}}
\newcommand{\bfPhi}{\bm{\Phi}}

\renewcommand\[{\begin{equation}}
\renewcommand\]{\end{equation}}

\usepackage[T1]{fontenc}

\makeindex

\begin{document} 
	\begin{CJK*}{UTF8}{gbsn} % Use default fonts from CJK (see below)
		\title{Phonon induced rank-2 U(1) nematic liquid states}
		\author{Han Yan (闫寒)}
		\email{hy41@rice.edu}
		\affiliation{Rice Academy of Fellows, Rice University, Houston, TX 77005, USA}
		\affiliation{Department of Physics \& Astronomy, Rice University, Houston, TX 77005, USA}
		\author{Andriy H. Nevidomskyy}
		\affiliation{Department of Physics \& Astronomy, Rice University, Houston, TX 77005, USA}
		\date{\today}
\begin{abstract}
Fascinating new phases of matter can emerge from strong electron interactions in solids. In recent years, a new exotic class of many-body phases, described by generalized electromagnetism of symmetric rank-2 electric and magnetic fields and immobile charge excitations dubbed \textit{fractons}, has attracted wide attention. Besides having interesting properties in their own right, the models with generalized electromagnetism are also closely related to gapped fracton quantum orders, new phases of dipole-covering systems, as well as quantum information and quantum gravity. However, experimental realization of the rank-2 U(1) gauge theory is still absent and even known practical experimental routes are scarce. In this work, we propose a scheme of coupled optical phonons and nematic degrees of freedom, as well as several concrete experimental platforms for their realizations. We show that these systems can realize the electrostatics sector of the rank-2 U(1) gauge theory. A great advantage of the proposed scheme is that it requires only the basic ingredients of phonon and nematic physics, and hence may be applicable to a wide range of experimental realizations from liquid crystals to electron orbitals.
%
%We expect that this  work will provide guidance for the realization of rank-2 U(1) and fracton states of matter on a variety of platforms.
\end{abstract}
\maketitle
\end{CJK*}

%\tableofcontents

\begin{figure}[t]
\centering
\includegraphics[width=\columnwidth]{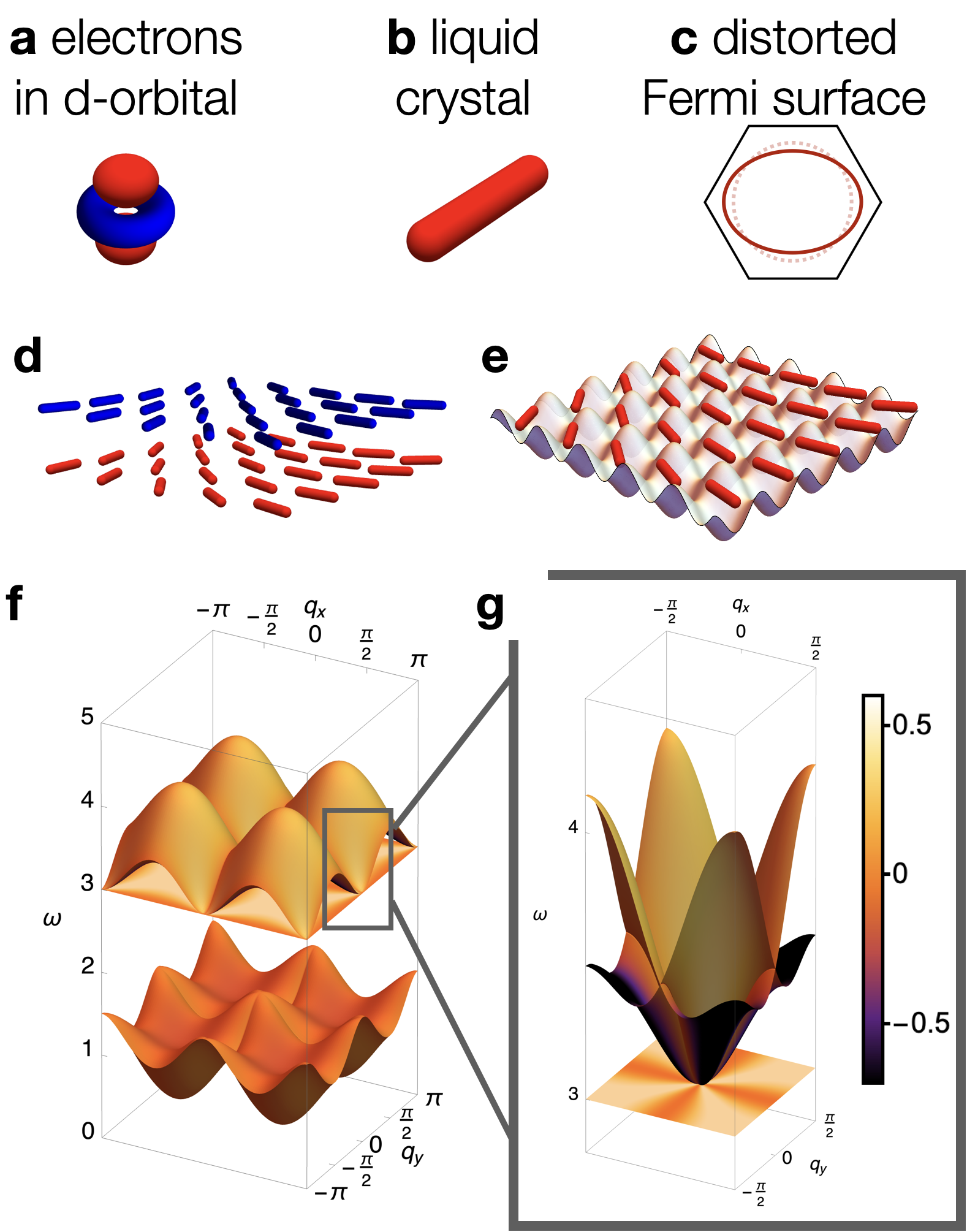}
\caption{
\textbf{Realizing rank-2 U(1) electrostatics via optical phonon-nematic coupling.}
\textbf{a-c} Examples of microscopic objects with nematic degrees of freedom. 
\textbf{d} Bilayer nematic as a representative experimental construction  that
realizes the ideal model [Eq.~\eqref{EQN_ideal_phonon_nam}].
\textbf{e} Nematic layer trapped in an artificial periodic potential as a representative experimental construction.
\textbf{f} The band structure of [Eq.~\eqref{EQN_ideal_phonon_nam}]. 
\textbf{g} Zoomed-in view of the phonon-nematic band structure,
in which the flat band corresponds to the vector-charge-free nematic configurations, 
and the upward dispersing bands 
correspond to the charged nematic configurations.
} 
\label{Fig_overview}
\end{figure}

\section{Introduction}

%``Emergence'' -- laws of nature arising as the effective theory describing many-body systems -- 
%is one of their most fascinating aspects \cite{anderson72}, both for intellectual curiosity and practical utility.
%
At the forefront of modern physics lies the concept of emergence -- the spontaneous appearance of qualitative changes in the behavior of large, complex systems that can by no means be inferred by extrapolating the properties of only a few particles~\cite{anderson72}. The emergent behavior is codified by the new `laws' in an effective theory, and the emergent phases often transcend the traditional Landau--Ginzburg paradigm of symmetry breaking. 
%In the past few decades, tremendous progress has been made in understanding phases beyond the traditional Landau-Ginzburg paradigm of symmetry breaking. 
%
One example of such an emergent phase are the spin liquids -- exotic states  built  on quantum superposition of  product states,
characterized not by 
any order parameter,
but by the topological entanglement
and topological order.
A subclass of such spin liquids can be described by local constraints on the local degrees of freedom (DoF), leading to the emergence of 
a   gauge invariant description, and thereby to topological orders,   
  fractionalised excitations and long-range entanglement 
\cite{Anderson1973,Balents2010,Savary2016,Zhou2017}. 
%
%For example, a fruitful field is to look for spin liquid states that are described by  local constraints  instead of order parameters.
%This can lead to the emergence of a local gauge symmetry, and thereby to topological orders, and exhibit fractionalised excitations and long-range entanglement  \cite{Anderson1973,Balents2010,Savary2016,Zhou2017}.
%
A well-known 
example is quantum spin ice
on the  pyrochlore lattice,
which realizes  $U(1)$ Maxwell gauge theory~\cite{hermele04PRB}.
It hosts
emergent excitations mimicking the Maxwell electrodynamics: photons, electric charges and even magnetic monopoles.
As such, it has been under intense 
theoretical  \cite{hermele04PRB,banerjee08-PRL100,benton12-PRB86,savary12,
Shannon2012PRL,hao14,Gingras14RoPP,kato15,chen17,huang18-PRL120,Balents2010,Savary2016,Zhou2017} 
and experimental 
\cite{zhou08,RossPRX11,fennell12,Kimura2013,sibille15,wen17,thompson17,Sibille2018,gao-arXiv} investigation.

%%%%%%%%%%%%%%%%%%%%%%%%%%%%%%%%%%%%%%%%%%%
% Paragraph 2 - "Recent work has highlighted..."
%%%%%%%%%%%%%%%%%%%%%%%%%%%%%%%%%%%%%%%%%%%
Recently, 
a class of more exotic forms 
of emergent electrodynamics  proposed as effective theories for spin liquid phases \cite{XuPRB06,PretkoPRB16,Rasmussen2016arXiv,PretkoPRB17}
has attracted considerable attention.
As a generalization of Maxwell electrodynamics,
it features 
electric and gauge  fields in the form of rank-2 (R2), or generally higher-rank, symmetry tensors.
The correspondingly modified Gauss's
conservation laws result in
some unexpected, exciting properties.
The electric charge excitations dubbed \textit{fractons} are 
intrinsically
constrained from moving
in the system,
and  foreshadow  a new class of gapped fracton quantum liquid order beyond topological order \cite{chamon05-PRL94,shannon04-PRB69,HaahPRA11,VijayPRB15,VijayPRB2016,BulmashPRB2018,MaPRB2018,Nandkishore2018arXiv,SlaglePRB17,GaborPRB17,SchmitzPRB2018,KubicaArXiv18}.
The rank-2 U(1)  (R2-U1) theories are
also shown to be akin to  gravity \cite{XuPRB06,Benton2016NatComm,PretkoPRD16,Yan2018arXiv},
and related to new phases of matter featuring dipole conserving dynamics   \cite{PretkoPRL18, gromov19,you2020fracton,zhou2021fractal,YZYou2021PhysRevB}.
However, these remarkable properties come with  a cost: the central ingredient --
local constraints applied to tensors -- is in a more complex form than the traditional Gauss's law of Maxwell electromagnetism.
To enforce these constraints,
complicated multi-body interactions are required in many
prototypical fracton models 
 \cite{chamon05-PRL94,XuPRB06,XuPhysRevD2010,XuPhysRevB2007,Rasmussen2016arXiv,HaahPRA11,VijayPRB15,VijayPRB2016}, 
while experimental proposals remain scarce \cite{SlaglePRB17,GaborPRL17,You2018arXiv}.
Therefore,
concrete designs for 
experimental realizations of R2-U1 phases pose a significant challenge, and overcoming this difficulty  would constitute a
crucial step for future development of the field.

Here we propose a realistic experimental scheme to 
achieve nematic liquid states
described by the classical limit R2-U1 theory, that is realizing 
 the electrostatics of such higher-rank theories.
Phases of matter with nematic DoF, such as   liquid crystals,
are good potential candidates for this purpose
since they
are naturally represented by symmetric tensors 
-- exactly those needed in the R2-U1   physics.
%
%Earlier works on elasticity-fracton duality have also exploited a similar 
%advantage, namely the elastic DOFs are also symmetric tensors \cite{Pretko2018PRL,gromov19PRL}.
The challenge is to find a realistic approach toward the specific low-energy Hamiltonian that would give rise to a nematic liquid state obeying the R2-U1 Gauss's law, 
instead of driving the system into an ordered state.

In this work we show that this is  {readily achievable}.
The ingredients in our model are  {quite common}:
Einstein phonons
and the most general coupling between phonons and nematic DoFs.\
%anc{I am wary of using the words "basic" or "simple" that appear to diminish our own contribution} 
%
{We demonstrate that} integrating out the phonon modes
leads precisely to the sought %\sout{vector-charged} 
Gauss's law-enforcing term on the remaining nematic DoFs.
Beside the idealized effective theory,
we present  a few 
concrete experimental platforms
where such a theory can be realized.
Our approach has the advantage 
of having a wide range of applicability.
The existence of nematics at different scales -- from electron orbitals 
to organic molecules, to soft matter --
means that our proposed design %\change{is not targeted  at a single system but a variety of them}
{can be realized in a variety of experimental platforms}.
Different types of nematic  matters available 
also enable us to construct
different versions of R2-U1 theories.
We hope that our work opens a gateway 
to experimental realizations of generalized higher-rank gauge theories.
%for a wide spectrum of the physics community.

%\section{The ideal model}
 
\section{The idealized model.}
\subsection{Hamiltonian of nematic-phonon coupling}
The ideal model Hamiltonian to realize the
R2-U1 physics via nematic-phonon coupling is composed of three parts: optical phonons, the nematic degrees of freedom, and their coupling: 
\[
\label{EQN_ideal_phonon_nam}
\begin{split}
\mathcal{H}_\text{classical} &  = 
\mathcal{H}_\text{ph} + \mathcal{H}_\text{ph-nem} + \mathcal{H}_\text{nem} 	\\
& =  \frac{\rho \omega_{0}^2}{2}  
\bm{u} \cdot \bm{u}   -\lambda  \varepsilon_{ij}\Phi_{ij}
+ M \sum_{i\le j }\Phi_{ij}^2,
\end{split}
%+ \mathcal{H}_\text{nem} 
\]
{where $\bm{u}(\bm{r})$ is the lattice distortion of the Einstein phonons (i.e., phonons with a flat energy dispersion $\hbar\omega_0$).}
%In $\mathcal{H}_\text{classical}$, the first term $\mathcal{H}_\text{ph}$
%is  for Einstein phonons (i.e., phonons on a flat band at energy level $\rho \omega_{0}^2/2$ without dispersion), where $\bm{u}(\bm{r})$ is the lattice distortion. 
The second term $\mathcal{H}_\text{ph-nem}$
is the   symmetric leading order coupling between 
the strain tensor of the lattice distortion
\[
\varepsilon_{ij} (\bfr) = \partial_i u_j(\bfr) + \partial_j u_i(\bfr),
\]
and the nematic DoF described by the symmetric tensor $\Phi_{ij}$ \cite{Cowley1976PhysRevB,Karahasanovic2016PhysRevB,Paul2017PhysRevLett,Carvalho2019PhysRevB,Fernandes2020}.
%The strain tensor is
%\[
%\varepsilon_{ij} (\bfr) = \partial_i u_j(\bfr) + \partial_j u_i(\bfr)	,
%\]
%and $\Phi_{ij}$ are the nematic DoF as a symmetric tensor.
The third term   $\mathcal{H}_\text{nem}$ has the meaning of a mass term and is assumed to be positive-definite ($M>0$) and can be thought of as imposing a physical constraint on the tensor to be of finite length. 
This occurs naturally in certain microscopic  nematic matters (see section \ref{sec:discussion} for details). 
%This constraint does not affect our key conclusions. 
In this work, we explicitly assume no spontaneous breaking of the rotational symmetry, i.e. we always assume nematic fluctuations without the long-range nematic order.~\footnote{Otherwise (if mass $M<0$) one would need to include quartic terms of the type $[\mathrm{Tr}(\Phi^2)]^2$ to stabilize the theory.}
%
%\an{\sout{The third term $\mathcal{H}_\text{nem}$
%is a simple on-site potential (mass term) for the nematics.
%Often
%the nematics are constrained by a ``unit-tensor''-like condition $\sum_{i\le j }\Phi_{ij}^2 = 1$ (see section \textbf{Discussion}  on their microscopic origins), so the third term can be dropped as long as the constraint is taken into account properly.}}
%
Finally, in this Hamiltonian we have suppressed the dynamical terms 
\[
\label{EQN_ideal_phonon_nam_dynamics}
\mathcal{H}_\text{dynamics} = (\partial_t \bfu)^2 + (\partial_t \bfPhi)^2,
\]
since we are mostly interested in the classical sector of the system.
Their potential role in a quantum system is also discussed in the section~\ref{sec:discussion} below.

The spectrum of the diagonized Hamiltonian in a square lattice is shown on Fig.~\ref{Fig_overview}(f,g) (see  Fig.~\ref{Fig_square_lattice} for the square lattice set up). The false color indicates the distribution on each band of the correlator $\braket{\Phi_{xx}(-\bfq)\Phi_{yy}(\bfq)}$, whose meaning will be clarified in a later part of this section. 
%The spectrum intensity distribution on each band is measured by $\braket{\Phi_{xx}(-\bfq)\Phi_{yy}(\bfq)}$,whose meaning will be clarified in a later part of this section. 

\begin{figure*}[t!]
\centering
\includegraphics[width=0.90\textwidth]{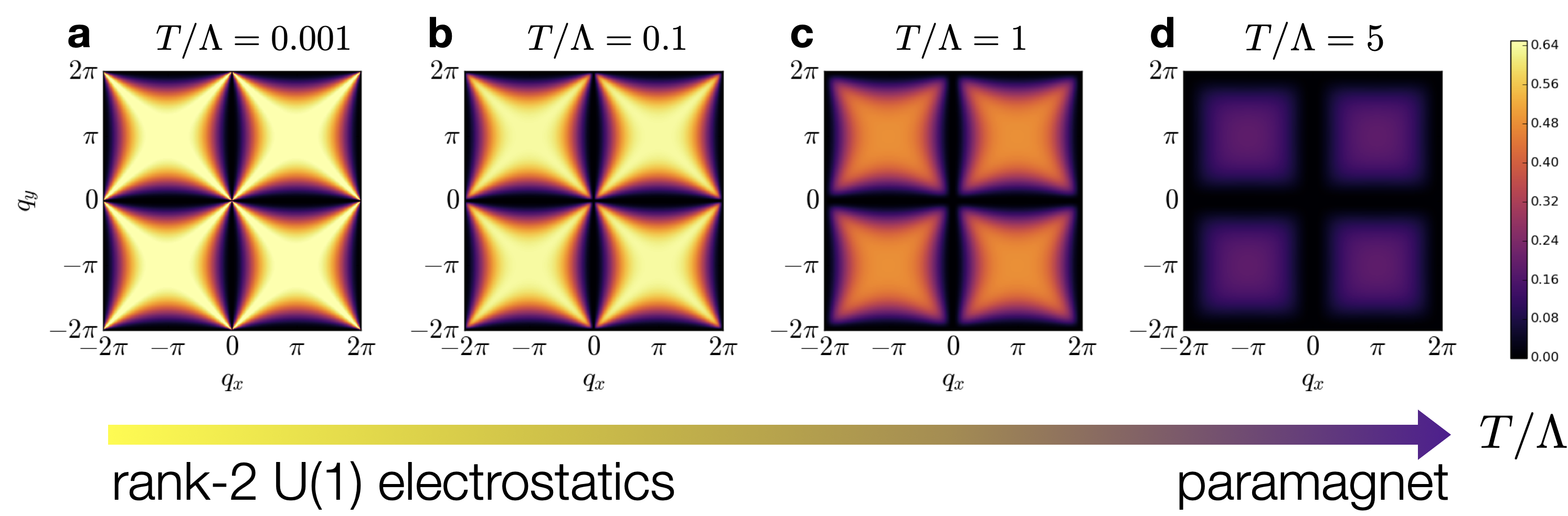}
\caption{\textbf{Nematic correlation functions $\langle \Phi_{xx} (-\bfq) \Phi_{yy}(\bfq) \rangle$ for model in Eq.~\eqref{EQN_nem_eff_r_space_ideal}.} The four panels show the correlation functions, using a false-color map, computed via the Self-Consistent Gaussian Approximation (see \textbf{Methods}) on a square lattice [Fig.~\ref{Fig_square_lattice}] at different temperatures relative to the parameter $\Lambda$.
The high temperature regime shown in \textbf{d} is a paramagnetic phase with vanishing correlations. 
The low temperature regime shown in \textbf{a} is the rank-2 U(1) phase, manifested by the 
characteristic 4-fold pinch point pattern
in the correlation function around  $\bfq = \bf{0}$, originating from  the functional form  $\langle \Phi_{xx} (-\bfq) \Phi_{yy}(\bfq) \rangle\propto q_x^2q_y^2/q^4$. \textbf{b,c} The 4-fold pinch points become gradually smeared due to thermal fluctuations at intermediate temperatures.
%The correlation function is computed by Self-Consistent Gaussian Approximatin.
} 
\label{Fig_SCGA}
\end{figure*}

%\begin{figure}[ht!]
%\centering
%\includegraphics[width=0.90\columnwidth]{asdf.png}
%\caption{\textbf{Heat capacity.}  
%The correlation function is computed by Self-Consistent Gaussian Approximatin.
%} 
%\label{Fig_HeatCapacity}
%\end{figure}

By integrating out the Gaussian phonon modes, we end up with the effective  theory
for the nematic DoF only, described by the Hamiltonian
%
%In momentum space, it is 
%
%let us examine the following ...
%
%\[
%\mathcal{H}_\text{nem-eff} = 
%\frac{\lambda^2}{2\rho\omega_{0}^2}  \Phi_{ij} \hat{q}_i \hat{q}_k \Phi_{kj}	+M \sum_{i\le j }\Phi_{ij}^2	.
%\label{EQN_nem_eff_q_space_ideal}
%\]
% 
\[
\mathcal{H}_\text{nem-eff} = 
%\frac{\lambda^2}{2\rho\omega_{0}^2}  
\Lambda (\partial_i \Phi_{ij}) (\partial_k \Phi_{kj}) +M \sum_{i\le j }\Phi_{ij}^2,
\label{EQN_nem_eff_r_space_ideal}
\]
where $\Lambda = {\lambda^2}/{(2\rho\omega_{0}^2)}$.
%
%In the limit of $\Lambda \equiv {\lambda^2}/{(2\rho\omega_{0}^2)}\gg T$,
{In the limit of sufficiently large $\Lambda \gg T$ relative to the temperature},
the first term imposes high energy cost for $\Phi_{ij}$ configurations that violate 
the constraint 
\[
\label{EQN_guass_law_nematic}
\partial_i \Phi_{ij}  = 0 .
\]
Upon identifying the nematic DoF with the generalized rank-2 electric field $\Phi_{ij} \longleftrightarrow E_{ij}$, its derivative becomes associated with the generalized vector charge:
\[
\label{EQN_nematic_electric_corr}
 \partial_i \Phi_{ij} \longleftrightarrow \rho_j \equiv \partial_i E_{ij} ,
\]
and the 
Eq.~\eqref{EQN_guass_law_nematic} becomes exactly the Gauss's law for the vector-charged R2-U1 theory.
Hence the classical R2-U1 nematic liquid state is realized in the low energy sector of the theory.

A more physical interpretation of the model is achieved by noticing that
\[
\label{EQN_phonon_vector_C_couple}
-\lambda \epsilon_{ij}\Phi_{ij}
= 2\lambda \bm{u}\cdot \bm{\rho} + \text{total derivative}.
\]
This means the vector charge excitation $\bm{\rho}$ is linearly coupled to the lattice distortion.
{The energy cost of the lattice distortion
induces, upon integrating out the lattice DoF, 
the potential energy ${\lambda^2}\bm{\rho}^2/{(2\rho\omega_0^2)} $ for the charge excitations.}

\subsection{Experimental signatures}

To quantitatively show the emergence of 
R2-U1 electrostatics, 
we study the model of Eq.~\eqref{EQN_nem_eff_r_space_ideal} on a square lattice
under the on-site constraint $\sum_{i\le j }\Phi_{ij}(\rr)^2 = 1$ and examine its correlation function $\langle \Phi_{ij} (-\bfq) \Phi_{kl}(\bfq) \rangle$ at different temperatures using the Self-Consistent Gaussian Approximation {(SCGA, described in Methods).}

The equal time correlation function $\braket{\Phi_{ij}(-\bfq)\Phi_{kl}(\bfq)}$ in the R2-U1 phase 
is constrained by the Gauss's law 
$
q_\alpha \langle \Phi_{ij} (-\bfq) \Phi_{kl}(\bfq) \rangle = 0,$ 
where $\alpha$ is one of the four indices $i,j,k,l$ and,
 the repeated index is summed over.
As a consequence the correlation is 
restricted to be proportional to  a highly anisotropic projector in the form of 
\begin{equation}
\label{EQN_4FPP_form}
\begin{split}
\langle \Phi_{ij} &(-\bfq) \Phi_{kl}(\bfq) \rangle  \propto \\
&  \frac{1}{2}\left(\delta_{i k} \delta_{j l}+\delta_{i l} \delta_{j k}\right)+\frac{q_{i} q_{j} q_{k} q_{l}}{q^{4}}\\
& -\frac{1}{2}\left(\delta_{i k} \frac{q_{j} q_{l}}{q^{2}}+\delta_{j k} \frac{q_{i} q_{l}}{q^{2}}+\delta_{i l} \frac{q_{j} q_{k}}{q^{2}}+\delta_{j l} \frac{q_{i} q_{k}}{q^{2}}\right) .
\end{split}
\end{equation}
In particular, 
$\langle \Phi_{xx} (-\bfq) \Phi_{yy}(\bfq) \rangle\propto q_x^2q_y^2/q^4$
shows a characteristic pattern dubbed ``4-fold pinch point'' \cite{PremPRB18,Yan2020PRL,benton2021arxiv}.

In Fig.~\ref{Fig_SCGA},
we present  the correlation function $\langle \Phi_{xx} (-\bfq) \Phi_{yy}(\bfq) \rangle$, computed within the SCGA approach,
at different ratios of $T/\Lambda$.
As the result demonstrates,
at high temperature, the system is  a paramagnet and the correlation function is essentially vanishing.
At low temperatures on the other hand, the 4-fold pinch point emerges as the system enters the R2-U1 phase. 
The transition between the two phases is expected to be not a phase transition but a crossover.

The same 4-fold pinch point is also visible in the band structure of the diagonlized Hamiltonian shown in Fig.~\ref{Fig_overview}(f,g).
There, the flat band corresponds to the nematic states obeying Gauss's law. 
{Note that the energy of this flat band is finite due to the non-zero mass $M$ in Eq.~\eqref{EQN_nem_eff_r_space_ideal}, and the lack of dispersion is due to the fact that the mass term imposes a local constraint, i.e. all such states are momentum independent.}
%It is at a finite energy due to the non-zero mass $M$ for nematics, but not dispersive, since all such states do not gain any additional energy.
The 4-fold pinch point is imprinted on the flat band, which is consistent with the result from the Gaussian-integrated theory [Eq.~\eqref{EQN_nem_eff_r_space_ideal}].

%\begin{figure}[ht!]
%\centering
%\includegraphics[width=0.90\textwidth]{}
%\caption{\textbf{Heat capacity.}  
%The correlation function is computed by Self-Consistent Gaussian Approximatin.
%} 
%\label{Fig_HeatCapacity}
%\end{figure}

\subsection{Effects of perturbations}
\label{subsec_perturbations}
The idealized model of Eq.~\eqref{EQN_ideal_phonon_nam} is  fine-tuned. 
For example, if there are symmetry-breaking terms, the nematic DoFs, instead of fluctuating subject to the Gauss's law, can become ordered at sufficiently low temperature (this would require setting mass $M<0$ and adding quartic terms to the Hamiltonian in Eq.~\eqref{EQN_ideal_phonon_nam}).
Hence, we must address the question whether the idealized model can be realized experimentally.

We remark that 
it should not be surprising that, generally speaking, the Hamiltonian realizing
such classical spin/nematic liquid  requires fine-tuning. 
A well known canonical example is the classical spin ice \cite{bramwell01},
whose exact macroscopic ground state degeneracy is the consequence of fine-tuned interactions, and can be lifted by addition of arbitrarily small terms to the Hamiltonian. The U(1) gauge theory being gapless, there is no protection against such terms in general. However, as long as these terms have a magnitude smaller than the experimentally accessible temperature, 
their effect is not strong enough to drive the system into e.g. an ordered state, and the relevant degrees of freedom fluctuate, subject to the constraint imposed by the Gauss's law.
This general principle applies equally to the spin ice and to our model of generalized higher-rank U(1) theory.

\section{Advantages and challenges of the idealized model.}
Several comments are in order before we continue with the 
discussion of the more concrete experimental  platforms to realize the idealized Hamiltonian in Eq.~\eqref{EQN_ideal_phonon_nam}.
First, 
this model has the advantage of being built upon rather common  elements:
the Einstein phonon is the zero-dispersion limit of an optical phonon, which is often a good approximation.
More generally, optical phonons with small dispersions also work,
since mild dispersion will only contribute to the higher-order terms.
Equally, the nematics DoF are common microscopic objects,
ranging in their origin from molecular anisotropy in classical liquid crystals, to orbital electron DoF in transition-metal compounds [cf. Fig.~\ref{Fig_overview}(a-c) and section~\ref{sec:discussion}]. The tendency towards the nematic distortion can also be emergent, for instance due to the Pomeranchuk instability of a Fermi surface \cite{Pomeranchuk}, discussed in more detail in section~\ref{sec:discussion}. %ranging from electrons in certain orbitals and distorted Fermi surfaces where quantum effects can be strong, to classical liquid crystals [cf. Fig.~\ref{Fig_overview}(a-c) and \textbf{Discussion}].
The phonon-nematic coupling in the second term of Eq.~\eqref{EQN_ideal_phonon_nam} is the lowest order coupling  that 
respects the rotational symmetry of the system and is also generally expected, as seen in many other studies \cite{Cowley1976PhysRevB,Karahasanovic2016PhysRevB,Paul2017PhysRevLett,Carvalho2019PhysRevB,Fernandes2020}.
%
%It is the lowest order coupling  that respecting rotational symmetry of the system.
%
Hence we   expect it to be the dominant term 
in relevant experiments.
%

%The remaining challenge of realizing Eq.~\eqref{EQN_ideal_phonon_nam}is that while all three elements are common in experiments, 
We emphasize that in previous theoretical studies,
the phonon-nematic coupling $\mathcal{H}_\text{ph-nem}$ 
was written for acoustic, rather than optical phonons, as discussed in detail in Refs.~\cite{Paul2017PhysRevLett,Carvalho2019PhysRevB,Fernandes2020}.
There, although
the coupling also yields a 4-fold  anisotropic susceptibility similar
to those shown in Fig.~\ref{Fig_SCGA},
the resulting effective theory is \textit{not} of the form of the sought-after rank-2 $U(1)$ electromagnetism.
%
%The gap of the phonons are crucial in deriving the Gauss's law-enforcing term in Eq.~\eqref{EQN_nem_eff_r_space_ideal}.
{The reason for demanding a finite (albeit possibly small) energy $\omega_0$ of optical phonons is to ensure that integrating out these higher-energy DoF is legitimate, leading to a finite $\Lambda \equiv {\lambda^2}/{(2\rho\omega_{0}^2)}$ in Eq.~\eqref{EQN_nem_eff_r_space_ideal}.}

In what follows we discuss 
concrete experimental set-ups 
that resolve the main challenge: 
%\change{coupling nematics to optical phonons  like the acoustic ones}
{how to implement the desired coupling between the nematic and optical-phonon degrees of freedom}. 

%\section{Proposals for optical phonon-nematic coupling}
\section{Experimental proposals}
\subsection{Bilayer construction}
%\textbf{Bilayer construction --}
For two-dimensional systems,
one solution we propose
is to construct 
systems with multiple sublattice sites.
Here we  consider an example of 
coupling two   layers together,
with each hosting 
the common
acoustic phonon-nematic coupling  [Fig.~\ref{Fig_overview}(d)].
%\anc{The figure reference appears incorrect. Did you mean a different panel in Fig. \ref{Fig_overview}?}

Each single layer, in the most symmetric case, is described by the Hamiltonian 
\[
\mathcal{H}_\text{ac-ph-nem}  
 = 2  \rho v^2  
(\partial_i u^X_j)
(\partial_i u^X_j)
 -\lambda  \varepsilon_{ij}^X\Phi_{ij}^X +
 M \sum_{i\le j }(\Phi_{ij}^{X})^2.
%+ \mathcal{H}_\text{nem} 
\label{EQN_acoustic_phonon_nam}
\]
%\[ 
%\mathcal{H}_\text{ac-ph-nem}  =  \frac{\rho}{2}  
%\varepsilon_{ij}^X C^{ijkl} \varepsilon_{kl}^X   -\lambda  \varepsilon_{ij}^X \Phi_{ij}^X
%+ \mu \Phi_{ij}^X\Phi_{ij}^X 
%+ \mathcal{H}_\text{nem} 
%\label{EQN_acoustic_phonon_nam}
%\]
%
Here $X = \text{T}, \text{B}$ corresponds to the top and bottom layer,
%The term $C^{ijkl}$ is a tensor of coupling constants,
%constrained by the symmetry of the system.
%
%For simplicity, we analyze the most symmetric case $C^{ijkl} = v^2 \delta^{ik}\delta^{jl}$,
%so that the Hamiltonian simplifies to 
%
%
and the acoustic phonon modes have 
isotropic linear dispersion $\omega_{ac}=v q$ (again here the dynamical terms are omitted).

We then consider the two layers coupled by the following interaction:
\[
\label{EQN_inter_layer_couple}
\mathcal{H}_\text{int} = g\rho (\matr{u}^\text{T} -\matr{u}^\text{B} )^2.
\]
%\change{This comes}
{Such interaction appears naturally} from an inter-layers atomic potential for the lattice sites penalizing their deviation
%deviating from their equilibrated 
from the equilibrium positions.

\begin{figure}[t!]
	\centering
	\includegraphics[width=0.6\columnwidth]{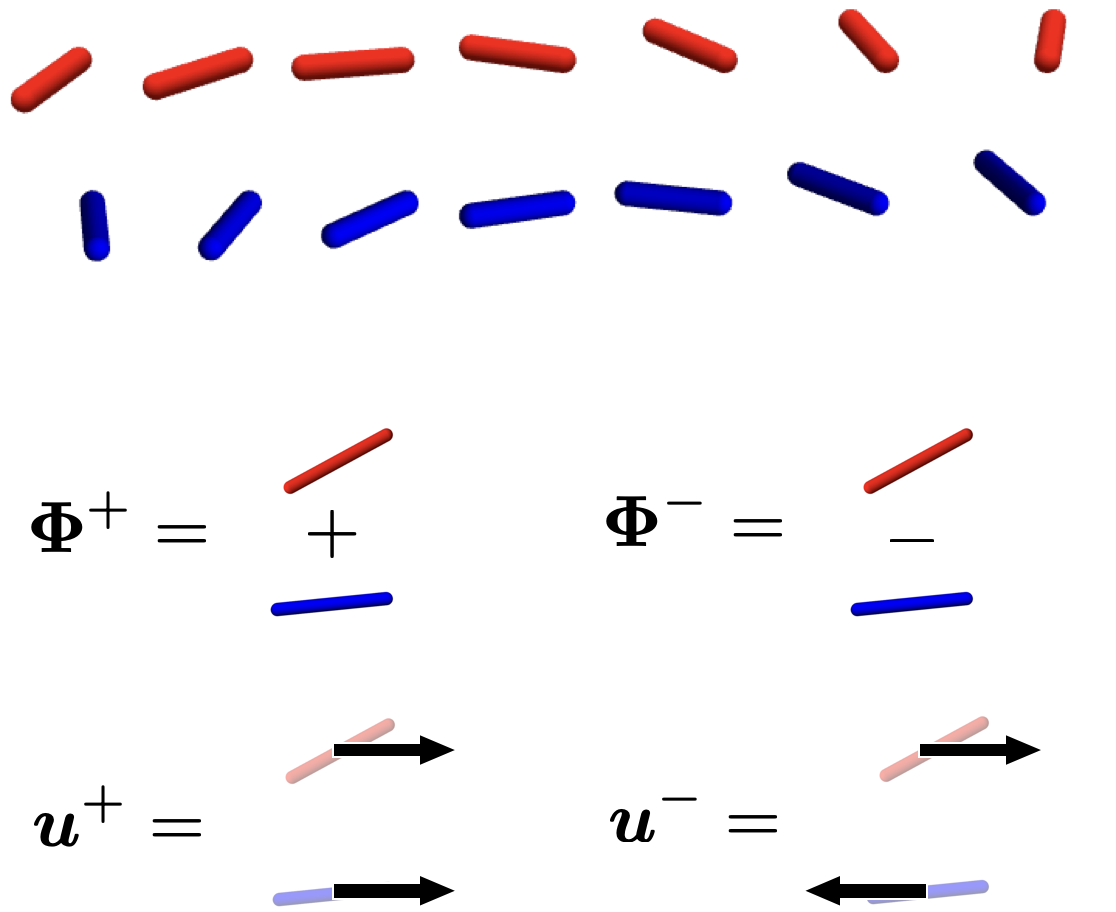}
	\caption{
	\textbf{A bilayer construction of two lattices with nematic degrees of freedom.} The atomic lattices are not shown for clarity but are essential in hosting intra-layer acoustic phonons. The inter-layer coupling results in the phonon splitting into two sectors: the acoustic in-phase mode ($u^+$) and the optical out-of-phase mode ($u^{-}$) in Eq.~\eqref{EQN_plus_minus_sec_def}. These two phonon modes couple to the corresponding nematic DoFs ($\Psi^+$ and $\Psi^{-}$) in the appropriate sectors. It is the coupling in the out-of-phase optical sector in Eq.~\eqref{EQN_ham_bilayer_minus} that leads to  the rank-2 U(1) theory.
	%Definition of the in-phase ($+$) and out-of-phase ($-$) sectors of phonon and nematic degrees of freedom in the bilayer construction [Eq.~\eqref{EQN_plus_minus_sec_def}].
	} 
	\label{FIG_bilayer}
\end{figure}

Diagonalizing $\mathcal{H}_\text{ac-ph-nem} +  \mathcal{H}_\text{int}$,
we find that the DoF can be decomposed into the in-phase and out-of-phase sectors labeled by $+,-$ [cf. Fig.~\ref{FIG_bilayer}],
\[
\label{EQN_plus_minus_sec_def}
\begin{split}
\bm{u}^\pm & = \frac{1}{\sqrt{2}}(\bm{u}^\text{T} \pm \bm{u}^\text{B}) , \\
\bm{\Phi}^\pm & = \frac{1}{\sqrt{2}}(\bm{\Phi}^\text{T} \pm \bm{\Phi}^\text{B} ).
\end{split}
\] 
%and they decouple from each other.
The two sectors decouple from each other.
The ``$+$''  sector  is described again by the usual acoustic phonon-nematic
coupling as in Eq.~\eqref{EQN_acoustic_phonon_nam}, and hence is not of our interest.
The out-of-phase ``$-$'' sector {describes the inter-layer optical phonon, coupled to the corresponding inter-layer nematic DoF}:
%\anc{Han, I think you were missing a factor of $v^2$ in front of the first term, which I've corrected. Please double-check:}
%has Hamiltonian
\[\label{EQN_ham_bilayer_minus}
\mathcal{H}_-
= 2\rho v^2 \sum_i (\bm{\nabla} u^-_i)^2 + 
2g  \rho\bm{u}^-\cdot \bm{u}^{-} 
-\lambda  \varepsilon^-_{ij}\Phi^-_{ij}
+ M \sum_{i\le j }\Phi^-_{ij}{}^2.
\]
Here, the phonons associated with $\bm{u}^-$
becomes gapped because of the inter-layer coupling [Eq.~\eqref{EQN_inter_layer_couple}].
The last three terms in Eq.~\eqref{EQN_ham_bilayer_minus}
are exactly what we are after.
The first term induces the
dispersion to the inter-layer optical phonon.
%For this term, 
Integrating out the photons, this term yields an additional, $\bf{q}$-depenent contribution of the order of
$\mathcal{O}\left(\frac{\lambda^2  v^2  q^4}{\rho g^2}(\bfPhi)^2\right)$. {This is to be compared to the principal term in the Gauss's law, of the order of $\mathcal{O}\left(\frac{\lambda^2 q^2}{\rho g}(\bfPhi)^2 \right)$.
}
Hence, if the dispersion scale is small compare to the gap, i.e.,
%\change{$v \ll g$}
{$vq_0^2 \ll g$} ($q_0 \sim 1/a$ denoting the edge of the Brillouin zone), then the phonon bands will be sufficiently flat,
and we obtain the idealized model of Eq.~\eqref{EQN_ideal_phonon_nam}
to a good approximation, {with $\Lambda = \frac{\lambda^2}{\rho g}$.}\\

%

%\textbf{Multiple sublattice site construction -- } 
\section{Mutliple sublattice sites}
\begin{figure}[ht]
	\centering
	\includegraphics[width=0.7\columnwidth]{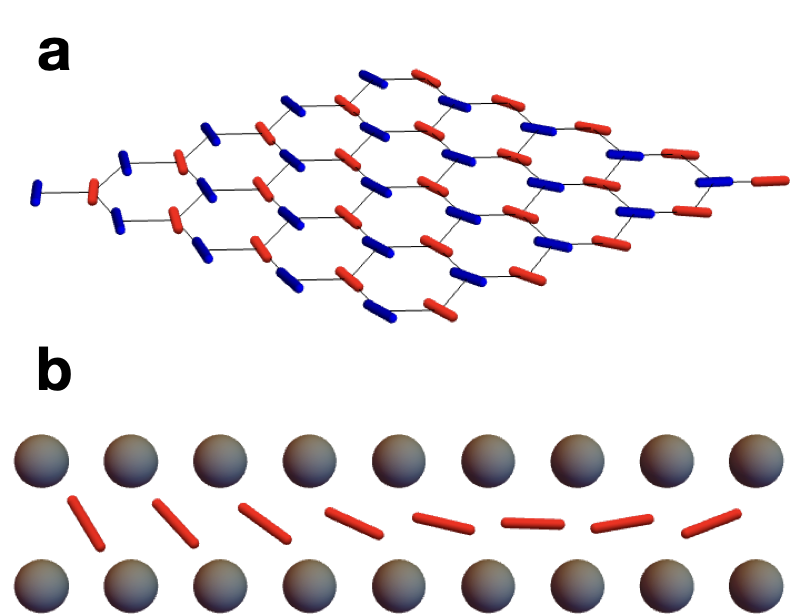}
	\caption{
	\textbf{Proposed experimental setups to realize the ideal model of Eq.~\eqref{EQN_ideal_phonon_nam}.}
	\textbf{a} The multiple sublattice site construction.
	Shown as an example is a 
	hexagon lattice of nematic degrees of freedom residing on two sublattice sites (blue and red rods).
	\textbf{b} The artificial potential well construction.
	The nematic layer (red rods) is sandwiched between two substrates of heavy molecules (grey balls). The substrates serve  as sources of the artificial potential term [Eq.~\eqref{EQM_potential_lattice}] for the lattice distortion $\bfu$ in the nematic layer.
	} 
	\label{Fig_2_sublattice}
\end{figure}

The essence of 
the  proposal in the previous section is that,
when there are multiple sublattice
sites in the system, the total number of phonons  increases accordingly, yet
only one set of them is acoustic, and the remaining phonon branches will become gapped, as desired to obtain the idealized model in Eq.~\eqref{EQN_ideal_phonon_nam}.
%Yet the coupling between phonons and nematics is preserved when the phonons become gapped.
%
Similar approaches can be designed following this principle.
For example, a single-layer nematic lattice with two sub-lattice sites per unit cell can also work [Fig.~\ref{Fig_2_sublattice}(a)].

%\textbf{Artificial potential well -- } 
\subsection{Artificial potential well}
Another scheme
we propose is 
to introduce an artificial potential 
for the 
%\change{nematic objects}
{nematic-site lattice displacements}, 
in order to break translational invariance and gap the phonons directly. 
%\anc{Han: are sure you mean the potential for nematic objects? The below equation is the potential for the lattice displacements $u$ instead...}
%
That is, 
we add a potential term 
\[
\label{EQM_potential_lattice}
\mathcal{H}_\text{pot} = \frac{\rho \omega_0^2}{2}\bfu\cdot \bfu
\]
to the lattice distortion, 
thus approximating the idealized model in Eq.~\eqref{EQN_ideal_phonon_nam}
when the phonon dispersion is mild.

The first realization of this idea is schematically illustrated in Fig.~\ref{Fig_overview}(e), wherein
the nematic atoms/molecules are placed in a periodic optical (laser) potential.
Such periodic potential
is a sophisticated experimental technique in use already \cite{Grimm2000,Bloch2005,Greiner2002,Bakr2009,Yang2020,MacDonald2003,RevModPhys.80.885,RevModPhys.89.011004}.

Another possible realization is to sandwich the nematic layer between the substrate layers of heavy molecules. The latter would then introduce a potential term 
to the nematics layer, as illustrated in Fig.~\ref{Fig_2_sublattice}(b).
%
%To summarise,
%one can start with the more conventional acoustic phonon-nematic coupled
%system. 
%By introducing coupling 
%between lattice sites of two such layers,
%the phonons get mixed and one of them become gapped. 
%In this way we have engineered the desired Einstein phonon-nematic coupling.
%
%In a similar spirit, 
%lattice with two or more sub-lattice sites, on which each sub-lattice site sits a nematic d.o.f., can also have optical phonon-nematic coupling from the same mechanism.
%
%
\begin{figure}[ht]
	\centering
	\includegraphics[width=0.75\columnwidth]{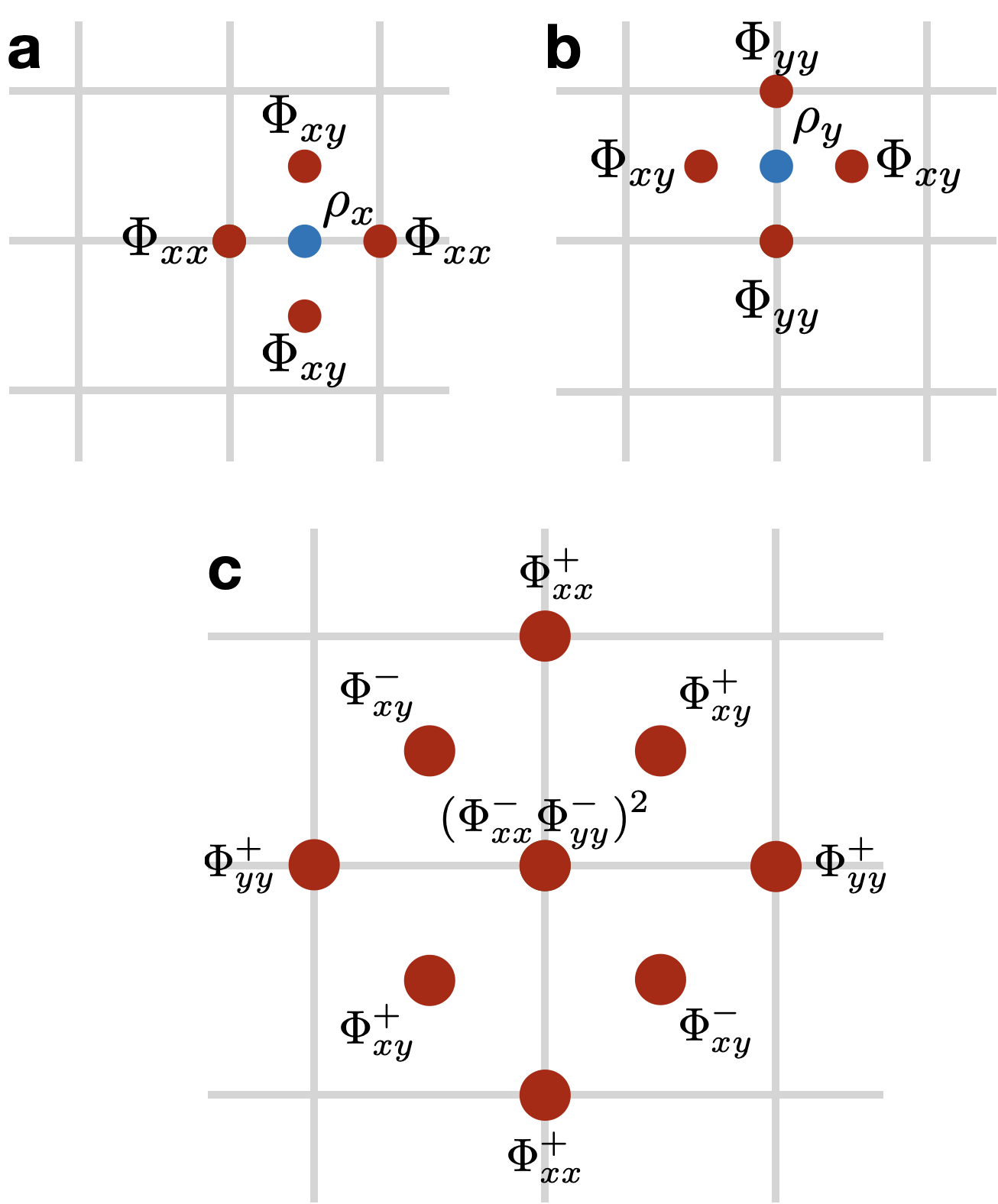}
	\caption{
	\textbf{The square lattice model of the nematic degrees of freedom.} This model is used for computing the correlation functions in Eq.~\eqref{EQN_4FPP_form} and throughout the paper.
	\textbf{a, b} Nematic degrees of freedom 
	$\Phi_{xx}$ and
	 $\Phi_{yy}$ defined
	on the vertices of the lattice, with off-diagonal components $\Phi_{xy}$ situated in the centers of the plaquettes.
	The lattice representation of the vector charge $\bm{\rho}$ in Eq.~\eqref{EQN_nematic_electric_corr} has two components $\bm{\rho} = (\rho_x, \rho_y)$ which live on the $x$- and $y$-links of the lattice, respectively, based on the four $\Phi_{ij}$ surrounding it.
%	 is defined as a lattice version of  Eq.~\eqref{EQN_nematic_electric_corr}	 based on the four $\Phi_{ij}$ surrounding it.
	\textbf{c}
	A dynamical term that acts as the gauge-invariant magnetic field in the generalized rank-2 electrodynamics. This term is a product of twelve $\phi_{ij}^\pm$ operators shown.
	When acting on a Gauss's law-obeying charge-free electric field configuration, the state is mapped onto another charge-free configuration.
	} 
	\label{Fig_square_lattice}
\end{figure}
%
%
%\section{Discussion} 
\section{Beyond the classical model.}
%\textbf{Quantum dynamics -- }
In this work we focused on how 
to achieve the electrostatics sector of the  rank-2 $U(1)$ theory.
This is a crucial step  toward the generalized quantum electrodynamics,
just as how the classical spin ice~\cite{bramwell01}
%\change{has been studied thoroughly before} 
provides the underpinnings for the development of a quantum spin ice~\cite{hermele04PRB,Gingras14RoPP}.

%Our main focus of this paper is  on the classical (electrostatics) sector of the nematic liquid state.
We now briefly discuss how quantum dynamics can arise in our classical model, which will render it the full-fledged rank-2 electrodynamics.
We start with a concrete example, and then discuss the general principles %\change{applied to all different lattices}
{applicable to all the implementations proposed above}. 

For concreteness, let us consider $\bfPhi$ living on the square lattice. For better visualization, 
we place the $\Phi_{xx}$, $\Phi_{yy}$ components on the vertices, and shift $\Phi_{xy} = \Phi_{yx}$ to the centers of the plaquettes.
This is illustrated in Fig.~\ref{Fig_square_lattice}(a,b). 
The generalized vector charges $\bm{\rho} = (\rho_x, \rho_y)$ are then defined on the links of the lattice. 
Specifically, $\rho_x$ is defined on the $x$-oriented links as 
$
\rho_x = \Delta_x \Phi_{xx} + \Delta_y \Phi_{yx},$
where $\Delta_i$ is the lattice derivative. 
Similarly $\rho_y$ is defined on $y$-links as
$
\label{EQN_rhoy}
\rho_y = \Delta_y \Phi_{yy} + \Delta_x \Phi_{xy}.
$
The classical sector of the Hamiltonian is 
\[
\mathcal{H}_\text{sq-cl}
= U \bm{\rho}^2 + M \sum_{i\le j }\Phi_{ij}^2.
\]

To introduce quantum dynamics, we argue by the way of analogy that each component of the tensor $\bfPhi$ %\change{is a quantum object like a large spin}
could be though of as corresponding to the $S^z$-component of a quantum spin, 
%with large spin number, 
and there is 
a generalized ``transverse field'' applied to the nematic DoFs,
\[
\mathcal{H}_\text{sq-dy} = h \sum_{i\le j }(\Phi_{ij}^+ + \Phi_{ij}^-),
\]
where $\Phi_{ij}^\pm$ are the raising and lowering operators of $\Phi_{ij}$.
Crudely speaking, $\Phi_{ij}^\pm$ plays the role of the gauge field operator $\bm{A}$ associated with the charge creation terms,
since they are %\change{dynamical}
{canonically} conjugate to the electric field components $\bm{E}$,
and creates charges when applied to an eigenstate of $\Phi_{ij}$.

A single operation of $\Phi_{xx}^\pm$
or $\Phi_{yy}^\pm$
will create  charges
in the system.
Within the sub-Hilbert space of the Gauss's law obeying states,
operators $\Phi_{ij}^\pm$
can only act on the Hilbert space  at a higher  perturbative order, such as to cancel all the charges created. 
An example we denote as $\Phi^+_\text{comp}$ is shown in Fig.~\ref{Fig_square_lattice}(c). 
There, a specific product of twelve  $\Phi_{ij}^\pm$ operators
connects one charge-free electric field configuration to another.
The fact that no charge is created anywhere  in the system
is equivalent to the statement that this composite product  of 12 operators 
is gauge invariant --
that is, $\Phi^+_\text{comp}$ (and also its hermitian conjugate $\Phi^-_\text{comp}$) plays  the role of the generalized magnetic field of the R2-U1 theory.
The generalized rank-2 electrodynamics is realized by the Hamiltonian
\[
\mathcal{H}_\text{sq-full}
= U \bm{\rho}^2 + M \sum_{i\le j }\Phi_{ij}^2 + \mu(\Phi^-_\text{comp}+\Phi^+_\text{comp}).
\]

Now let us comment on the general properties of the quantum dynamics of the nematic R2-U1 theory.
Like in a quantum spin ice,
the emergent magnetic field usually involves
multiple operators, and is generated perturbatively via the 
%\change{transverse field  and neighbouring spin coupling terms}
{product of transverse field operators}
which preserve the Gauss's law.
%which form a charge-free composite operation. 
%
In the conventional Maxwell U(1) theory, 
these composite operators
are simply loops of the dynamical operators, forming a lattice realization of the magnetic flux $\oint \nabla \times \bm{A}\, \ud l = \iint\bm{B}\, \ud \sigma$.
In R2-U1 theory,
the composite operators become more complicated as shown in the square lattice example above.

Although the long wavelength theory will remain the same,  
the available quantum dynamical terms will
depend on the details of the lattice geometry  and the microscopic implementation of the nematic DoF.
It is also possible that the quantum dynamics leads the system into other, ordered phases instead of R2-U1 electrodynamics (this is true of the quantum spin-ice as well).
The exact consequences will have to be discussed on a case-by-case basis.
%

%\textbf{Quantum dynamics.}

%
%They introduce dynamics to the propagation of electric field,
%which become photons.

%The physics achieved in this paper is classical, without the magnetic field part, equivalent to 
%Eq.~\eqref{EQN_Classial_R2U1}.

%Generally speaking, if there are some  dynamical  terms for the nematics -- for example a ``transversed field'', or a term  $\Dot{\bfPhi}^2$ -- 
%then at higher perturbative order they may be combined to  act as tunneling terms
%within the Hilbert space of charge-free electric field. 

%That is, the dynamics of $\bfPhi$ plays a role similar to $\bf{A}$, 
%as the electric field and gauge field are conjugate to each other. 
%Under the Gauss's law constraint, 
%the $\bf{A}$ form composites that change the configuration of $\bfPhi$ without creating charges in the system, which is essentially the gauge-invariant $\bf{B}^2$ term.

%Since this paper aims to outline a generic experimental route toward
%R2-U1 physics, we will not discuss such specifics.\\
 
\section{Discussion: Microscopic origin of the nematics.}\label{sec:discussion}
%\textbf{Microscopic origin of nematics -- }
In our construction,
we tacitly assumed
that the nematic DoF are described by a symmetric tensor with all its independent components, of which there are 3 in the two-dimensional systems and 5 in three-dimensional ones.
Depending on the microscopic origin of the nematics,
the number of DoF in the symmetric tensor representation may be fewer than those numbers.
%\sout{, or discretized.}\anc{I know what you mean, Han, but it may be better to explain it in a seperate paragraph, for now focusing only on the number of independent components.}
Below, we provide several concrete examples of the various microscopic realizations of the nematic DoFs.

The first example is that of $d$-electrons in transition metals (Fig.~\ref{Fig_overview}(a)).
There are in total five such orbitals corresponding to the $|l=2; m=-2,-1,\dots ,2\rangle$ states in the spherical harmonic expansion.
They form a symmetric, traceless tensor representation of the group $SU(2)$ describing rotations in the orbital Hilbert space (the $l=0,1$ representations are the trace and anti-symmetric components of this matrix). Ignoring the crystal field effects, which generically lift the orbital $SU(2)$ symmetry, these five orbitals form a degenerate manifold, out of which an orbital-nematic order can appear if the symmetry is spontaneously broken. In the disordered, symmetry-preserving phase, these orbital degrees of freedom can be used to construct classical rank-2 electrostatics as outlined in this work.
Moreover, by virtue of being intrinsically quantum objects, such models are also good candidates for constructing quantum electrodynamics of R2-U1 theory.

The second example of the nematics is a classical liquid crystal. In 2D, such as shown schematically in Fig.~\ref{Fig_overview}(b), the nematicity is described by
a director of a fixed length, encoded in a $2\times2$ symmetric matrix  
\[
\bfPhi
=
\begin{pmatrix}
\cos2\theta & \sin2\theta \\
\sin2\theta & - \cos2\theta
\end{pmatrix}.
\label{eq.theta-nematic}
\]
{Note that the matrix is traceless and unimodular (reflecting the fact that director is of unit length), and as a result, nematic DoF are described not by three but by a  single independent parameter, the azimuthal angle $\theta$.  The idealized theory presented in the beginning still holds, however the lack of the necessary rank-2 DoFs means the proper R2-U1 electrostatics cannot be realized.

{Another example of the nematicity is the spontaneous distortion of the Fermi surface (see e.g. Fig.~\ref{Fig_overview}(c)), known as the Pomeranchuk instability\cite{Pomeranchuk}, which in the simplest case of an isotropic  (circular in 2D) Fermi surface is described by the quadrupole density operator (see e.g. Ref.~\onlinecite{Oganesyan})
\beq
\bfPhi_{FS}(\qq) = \frac{1}{k_F^2} \psi^\dagger(\qq)
\begin{pmatrix}
q_x^2 - q_y^2 &  q_xq_y \\
q_x q_y & q_y^2 - q_x^2
\end{pmatrix}
\psi(\qq),
\eeq
where $\psi^\dagger(\qq)$ and $\psi(\qq)$ are the electron creation/annihilation operators at momentum $\qq$. The above matrix is also traceless, yielding traceless R2-U1 theory upon integration of the phonon modes coupled to $\bfPhi_{FS}$ as in Eq.~\eqref{EQN_ideal_phonon_nam}. The elliptic Fermi surface distortion thus has two independent DoFs:   $\Phi_{xx}$ and $\Phi_{xy}$, which can also be cast in the form of a complex order parameter $Q e^{i2\theta}=\Phi_{xx} + i\Phi_{xy}$, with the amplitude $Q$ proportional to the eccentricity of the ellipse and angle $\pm\theta$ its azimuthal direction. 
}

{We note that in the above example, the presence of the underlying crystalline lattice can pin the Fermi surface distortion along particular direction(s), such as shown in Fig.~\ref{Fig_overview}(c).
For instance, pinning to $\pm x$ or $\pm y$ directions on the square lattice  introduces a potential $U_{lat}(\theta) = -U_0\cos(2\theta)$ for the azimuthal angle. The resulting rank-2 theory would then become discrete, described by a 4-state Potts model on a square lattice (rather than the continuous U(1) parameter). Nevertheless, for temperatures and energy scales above $U_0$, the classical theory could be approximately described as having a continuous $U(1)$ symmetry. 
}

{Our final example of the (discrete) nematic order is realized on crystalline lattices with $n$-fold irreducible representations ($n=2,3$) of the point group. For instance, hexagonal systems (with point groups $C_6$ and $D_6$ in 2D) allow two-dimensional irreducible representations and hence the nematic order parameter can be parametrized by $\bfPhi = \Phi_0(\cos(2\theta), \sin(2\theta))$, which can be cast in the form of a traceless rank-2 tensor as in Eq.~\eqref{eq.theta-nematic}. This well known fact has been exploited recently in the discussion of nematicity in the twisted bilayer graphene, where coupling to acoustic phonons (different from the optical phonons in our case) was also considered\cite{Fernandes2020}. 
Generically, the lattice pinning will result in a discrete Potts model description of the nematic DoF, analogous to the previous case, and upon integrating out the (optical) phonons, the resulting rank-2 theory will be a discrete one.}

When designing possible experimental realizations of the nematic-phonon coupling,
one should thus be aware of the consequence of such discretization and the decreased number of the DoF {(as exemplified by the traceless condition in example 2 and 3 above),}
since too few DoF may result in ordered phases or states with subsystem symmetries only.
This however could also be a blessing in disguise, since it means we have a 
wider range of R2--U1 theories accessible in an experiment.
A particularly interesting type of such theories,
for instance,
is built in 3D from tensors with all diagonal components vanishing.
Such ``hollow'' rank-2 theories turn out to be the gateways toward  gapped
fracton order uppon ``Higgsing" the rank-2 U(1) degrees of freedom \cite{BulmashPRB2018,MaPRB2018}. The resulting gapped fracton orders hold
a great potential for 
applications in quantum memory storage.\\

%A.N.: looking at published Nature Comm papers, the subheading "Summary" is unnecessary, a paragraph starting with "In summary..." is all that's needed:
%\textbf{Summary --}
In summary,
we presented a theoretical model with simple ingredients that can realize the emergent rank-2 $U(1)$ electrostatics via optical phonon-nematic coupling. {Given the intimate connection between this rank-2 generalized electrodynamics and the exotic fracton phases of matter~\cite{PretkoPRB16, PretkoPRB17} which have recently garnered much attention, the present work thus paves the way towards natural implementations of the fracton matter in the experiment.}
%Its simplicity means high applicability in experiments, and we outlined several such constructions.
{Given the simplicity of the ingredients (optical phonons and nematic DoF), we hope this proposal may be realized in various settings, from liquid crystals to bilayer systems, to polar molecules in a periodic optical potential, and we have outlined several such possible constructions.}
%% A.N. the following sentence repeats unnecessarily what has been stated a couple of sentences earlier.
%
%We anticipate our work to  provide crucial guidance to experimental realization of R2-U1 phases of matter for a wide variety of experimentalists.
%
{The present proposal yields a classical rank-2 theory, which is a necessary first step on the path towards truly quantum rank-2 electrodynamics and fracton physics. We have outlined a possible route towards such quantum theory by incorporating 
%``transverse'' quantum terms 
the dynamics of the generalized magnetic fields
into our nematic model.} 
%We have also discussed the path towards incorporating ``transverse'' quantum terms into the Hamiltonian, which will result in a truly quantum rank-2 electrodynamics.

\textit{Acknowledgements.} The authors thank Leo Radzihovsky for discussions. This work was supported by the National Science Foundation Division of Materials Research under the Award DMR-1917511.  
 
\renewcommand{\emph}{\textit}
\bibliography{reference} 
\appendix

\section{Brief Review of Rank--2 U(1) Gauge Theory}
We start by briefly reviewing a version of rank--2 U(1) gauge theory,
which is to be realized in the models we propose in this paper.

As its name suggested, the R2--U1 gauge theory uses rank--2 tensors
$E_{ij}$ and $A_{ij}$ as its electric and gauge field 
 instead of vectors.
More specifically,
the tensor field is symmetric,
\[
 E_{ij} = E_{ji}\ , \;  A_{ij} = A_{ji}\; .
\] 

The charge is a vector defined as
\[
\rho_i = \partial^kE_{ki}\	.
\]
The low-energy sector of the theory has to be charge-free,
\[
\partial^kE_{ki}= 0,
\]
which dictates the form of the gauge invariance condition
\[
A_{ij }\rightarrow A_{ij} + \partial_i\lambda_j +\partial_i \lambda_j\;.
\]
The magnetic field is the simplest object that is gauge-invariant,
\[
B_{ij} = \epsilon_{i a b} \epsilon_{j c d} \partial^{a} \partial^{c} A^{b d}	.
\]

One can now write down the Hamiltonian for the R2-U1 gauge theory as
\[
\begin{split}
\mathcal{H}_\text{R2-U1} &=  U\partial^kE_{ki} \partial^lE_{li} + E_{ij}E_{ij} + B_{ij} B_{ij}\\
&= U\bm{\rho}^2 + \bm{E}^2 +\bm{B}^2
\end{split}
\]
Here we assumed the Einstein's summation rule while not caring about the super- and sub-scripts.

Our aim in this paper is to find out a general, and experimentally realistic routes 
to realize the classical part of this Hamiltonian
\[\label{EQN_Classial_R2U1}
\mathcal{H}_{\text{R2-U1-cl}} = U\bm{\rho}^2 + \bm{E}^2 .
\]
The quantum dynamics, i.e. the $\bm{B}^2$ term, is also possible to
realize, but is highly dependent on the specific set up of the physical system.
It will not be a focus of this paper.

\section{Self-consistent Gaussian Approximation.} The Self-Consistent Gaussian Approximation  (SCGA)
is an analytical method that treats
the nematics in the large-$N$ limit,
which is known to produce rather accurate results in the spin/nematic liquid phases.
Our calculation follows closely the exposition in
Ref.~\onlinecite{Isakov2004PhysRevLett}.
We first treat $\Phi_{ij}$ as independent, freely fluctuating DoF.
The Hamiltonian in the momentum space is written as
\begin{equation}
\mathcal{E}_{\text{Large-}N} = % \frac{1}{2}\mathbf{S}
%\begin{pmatrix}
%	\mathcal{H}_{xx} & 0& 0  \\
%	0 &   \mathcal{H}_{yy} & 0 \\
%	0 & 0&  \mathcal{H}_{zz}  \\
%\end{pmatrix}
%\mathbf{S}^T
 \frac{1}{2}\tilde{\bm{\Phi}}\mathcal{H}_{\text{Large-}N}\tilde{\bm{\Phi}}^T ,
\end{equation}
written in terms of the triad of nematic components (for the two-dimenstional model)
%$\tilde{\bm{\Phi}} = (\Phi_{xx},\Phi_{xy},\Phi_{xy})$ 
%%A.N. I think there is a typo in the 2nd component:
$\tilde{\bm{\Phi}} = (\Phi_{xx},\Phi_{yy},\Phi_{xy})$. 
%in case of two-dimensional model.
The  matrix   ${H}_{\text{Large}-N}$
is the Fourier transformed interaction matrix from Eq.~\eqref{EQN_nem_eff_r_space_ideal}:
%It is explicitly written as 
\[
{H}_{\text{Large-}N} = 
2\Lambda\begin{pmatrix}
  C_x^2 & 0 &  C_x C_y\\
0 &   C_y^2 &  C_x C_y\\ 
 C_x C_y &  C_x C_y&   C_x^2+  C_y^2
\end{pmatrix} ,
\]
where $C_x$ and $C_y$ are the momentum dependent functions. For the square lattice model [Fig.~\ref{Fig_square_lattice}], \mbox{$C_x = 2\sin(q_x/2)$},
$C_y = 2\sin(q_y/2)$, with the lattice constant set to $1$.
%\begin{equation}
%\mathcal{H}_{xx}  =   2 J_x
%\begin{pmatrix}
%0 & \cos(\bfq\cdot \bfr_{12}) & \cos(\bfq\cdot \bfr_{13}) & \cos(\bfq\cdot \bfr_{14})  \\
%\cos(\bfq\cdot \bfr_{21})  & 0 & \cos(\bfq\cdot \bfr_{23}) & \cos(\bfq\cdot \bfr_{24} \\
%\cos(\bfq\cdot \bfr_{31}) & \cos(\bfq\cdot \bfr_{32}) & 0 & \cos(\bfq\cdot \bfr_{34}) \\
%\cos(\bfq\cdot \bfr_{41}) & \cos(\bfq\cdot \bfr_{42}) & \cos(\bfq\cdot \bfr_{43}) & 0 \\
%\end{pmatrix}
%,
%\end{equation}
%where $\bfr_{ij}$ is the relative position between sublattice sites $i, j$.
%Similarly    $\mathcal{H}_{yy}$ and 
%$\mathcal{H}_{zz}$ are defined by replacing $J_x$ with $J_y$ and $J_z$ respectively.

We then introduce a Lagrange multiplier
with coefficient $\mu(\beta)$ 
to the partition function
to obtain
\begin{equation} 
    \mathcal{Z} =   \exp\left( -\frac{1}{2} {\int_\text{BZ}  \text{d}\bfq \int\text{d}\tilde{\bm{\Phi}}   \tilde{\bm{\Phi}}  \left[\beta  \mathcal{H}_{\text{Large-}N}+ \mu(\beta)
	 \mathcal{I}\right]\tilde{\bm{\Phi}}^T }  \right),
\end{equation}
where $\beta$ denotes the inverse temperature.
The purpose of the term $\mu(\beta) \tilde{\bm{\Phi}} 
	 \mathcal{I}\tilde{\bm{\Phi}}^T $ ($\mathcal{I}$ stands for the identity matrix)
is to impose, on average, an additional unimodular constraint
on the nematic DoF, such that
\begin{equation}
  \langle  \Phi_{xx}^2  +  \Phi_{yy}^2 +   \Phi_{xy}^2 \rangle  = 1 .
\end{equation}
For a given temperature $T= 1/\beta$, the value of $\mu(\beta)$ is numerically  obtained by searching for its value that must satisfy the constraint
\begin{equation}
\int_\text{BZ}   \text{d} \bfq \sum_{i=1}^{3}\frac{1}{\lambda_i(\bfq)+\mu(\beta)} =  \langle  \Phi_{xx}^2  +  \Phi_{yy}^2 +   \Phi_{xy}^2 \rangle = 1   ,
\end{equation}
where $\lambda_i(\bfq),\ i=1,2,3$
are the three eigenvalues of $\beta\mathcal{H}_{\text{Large-}N}(\bfq)$.

With $\mu$ fixed,
the partition function 
is completely determined for a free theory
of $\tilde{\bm{\Phi}}$,
and all correlation functions in Fig.~\ref{Fig_SCGA} can be computed from extracting the corresponding components in  $\left[\beta  \mathcal{H}_\text{Large-N}+ \mu(\beta)
\mathcal{I}\right]^{-1}$.\\

%
%Thus, studying the dependence of the thermodynamic quantities, such as magnetization and specific heat, on the strength and direction of magnetic field allows us to determine the parameters of the Hamiltonian Eq.~\eqref{eq:Hnn} as well as the values of the $g$-factors in Eq.~\eqref{eq:g-factors}. To this end, we have performed the exact diagonalization (ED) calculations on the 16-site cluster and computed the magnetization as a function of applied field strength, shown in Fig.~\ref{fig:fig2}(a) and (b). The red, blue and green curves correspond to the different temperatures in the experiment, which are fitted very well using our model. 
%}\\
\end{document}